\documentclass[twocolumn,showpacs,preprintnumbers,amsmath,amssymb]{revtex4}

\usepackage{graphicx}
\usepackage{dcolumn}
\usepackage{bm}

\begin{document}


\title{Model for the occurrence of Fermi pockets without the pseudogap hypothesis 

in underdoped cuprate superconductors 

-  Interplay of Jahn-Teller physics and Mott physics -}

\author{Hiroshi Kamimura}
 \email{kamimura@rs.kagu.tus.ac.jp}
 \affiliation{%
Research Institute for Science and Technology, Tokyo University of Science, 1-3 Kagurazaka, Shinjuku-ku, Tokyo 162-8601, Japan
}%
\author{Hideki Ushio}%
 
\affiliation{%
Tokyo National College of Technology, 1220-2 Kunugida-chou, Hachioji 193-0997, Japan
}%

\date{\today}

\begin{abstract}
Central issues in the electronic structure of underdoped cuprate superconductors are to clarify the shape of the Fermi surfaces and the origin of a pseudogap. 
Based on the model proposed by Kamimura and Suwa which bears important characteristics born from the interplay of Jahn-Teller Physics and Mott Physics, we show that the feature of Fermi surfaces is the  Fermi pockets constructed by doped holes under the coexistence of a metallic state and of the local antiferromagnetic order. Below $T_{\rm c}$ the holes on Fermi pockets form Cooper pairs with d-wave symmetry in the nodal region. In the antinodal region all the states below the Fermi level are occupied by electrons so that there is no gap, not even pseudo both below and above $T_{\rm c}$. Calculated angle-resolved photoemission spectrum below $T_{\rm c}$ show a coherent peak at the nodal region while a broad hump in antinodal region. From this feature the origin of the two distinct gaps in observed ARPES is elucidated. The finite-size-effects of a spin-correlation length coexisting with a metallic state are discussed. In particular, we discuss a possibility of the spatially inhomogeneous distribution of Fermi-pocket-states and of large-Fermi-surface-states above $T_{\rm c}$ which changes with time. Finally a new phase diagram for underdoped cuprates is proposed.
\end{abstract}

\maketitle

\section{Introduction}
Undoped copper oxide La$_2$CuO$_4$ is an antiferromagnetic Mott insulator, in which an electron correlation plays an important role \cite{Anderson}. Thus we may say that undoped cuprates are governed by Mott physics. In 1986 Bednorz and M\"{u}ller discovered high temperature superconductivity in copper oxides by doping hole carriers into La$_2$CuO$_4$ \cite{Bednorz}. Their motivation was that higher $T_{\rm c}$ could be achieved for copper oxide materials by combining Jahn-Teller (JT) active Cu ions with the structural complexity of layer-type perovskite oxides. In order to investigate the mechanism of high temperature superconductivity, most of models assumed that the doped holes itinerate through the orbitals extended over a CuO$_{2}$ plane in the systems consisting of the CuO$_{6}$ octahedrons elongated by the JT effect. Those models are called ``single-component theory'', because the orbitals of hole carriers extend only over a CuO$_{2}$ plane.

In 1989, Kamimura and his coworkers showed by first-principles calculations that the apical oxygen in the CuO$_{6}$ octahedrons tend to approach towards Cu$^{2+}$ ions, when Sr$^{2+}$ ions are substituted for La$^{3+}$ ions in La$_2$CuO$_4$, in order to gain the attractive electrostatic energy in ionic crystals such as cuprates \cite{Shima,Oshiyama}. As a result the elongated CuO$_{6}$ by the JT effect shrink by doping holes. This deformation against the JT distortion is called ``anti-Jahn-Teller effect'' \cite{Copper Oxide}. By this effect the energy separation between the two kinds of orbital states which have been split originally by the JT effect becomes smaller by doping hole-carriers, where the spatial extension of one kind is parallel to the CuO$_{2}$ plane while that of the other kind is perpendicular to it. 

By taking account of the anti-Jahn-Teller effect Kamimura and Suwa proposed that one must consider these two kinds of orbital states equally in forming the metallic state of cuprates, and they constructed a metallic state coexisting with the local antiferromagnetic (AF) order \cite{KS}. This model is called ``Kamimura-Suwa (K-S) model'' \cite{Kamimura_Entropy}. Since ``anti-Jahn-Teller effect'' is a central issue of the Jahn-Teller physics, we may say that the K-S model bears important characteristics born from the interplay of Jahn-Teller Physics and Mott Physics. Since these two kinds of orbitals extend not only over the CuO$_{2}$ plane but also along the direction perpendicular to it,  the K-S model represents a prototype of ``two-component theory'', in the contrast to the single-component theory.

In this paper we will discuss three important subjects emerged from the interplay of Jahn-Teller Physics and Mott Physics, based on the K-S model. The first one is concerned with the shape of Fermi surfaces in underdoped cuprates. The clarification of Fermi surface structures is nowadays a central issue in underdoped cuprates. There are two views for Fermi surfaces in cuprates. One view is based on the single-component theory, where a metallic state has a large Fermi surface (FS) \cite{Marshall, Norman_Nature, Kanigel}. Since angle-resolved photoemission spectroscopy (ARPES) experiments did not show the evidence of a large FS, the phenomenological idea of a pseudogap was introduced \cite{Norman_PRB2007}. An alternative view is based on the two-component theory developed by Kamimura and Suwa \cite{KS, KU, UK, Kamimura_Entropy}, and they have shown that the coexistence of a metallic state and local antiferromagnetic(AF) order results in the Fermi pockets constructed by doped holes in the nodal region. This key-point of the K-S model results in the coexistence of superconductivity and the local AF order below $T_{\rm c}$, as shown by Kamimura and coworkers. \cite{Kamimura_d-Wave, Kamimura_Super}. The appearance of Fermi pockets and small Fermi surfaces in cuprates has been recently reported by various experimental groups \cite{Meng, Nicolas Doiron-Leyrud, Bangura, Yoshida1, Yoshida2, Charkravarty_Kee}.

The second one is concerned with the phenomenological concept of pseudogap and the clarification of observed  two-gaps-structure in ARPES spectra in the underdoped regime. Based on the electronic structure calculated from the K-S model \cite{KU, UK}, we will show that the  angle-resolved photoemission spectroscopy (ARPES) 
spectra in cuprates below $T_{\rm c}$ exhibit a peculiar feature consisting of a coherent peak 
due to the superconducting density of states at the nodal region and the real transitions of 
electrons from the occupied states below the Fermi level to a free-electron 
state above the vacuum level in the antinodal region. Since the latter transitions 
reflect the density of states (DOS) of the highest energy band for the doped 
holes, these transitions may appear as a broad hump with a small peak. In this context it will be concluded that the pseudogap in the underdoped regime is absent, based on the K-S model.

Concerning the ARPES experiments in underdoped cuprates, Tanaka and his coworkers reported very interesting gap features in the observation of ARPES spectra. Their result exhibits a coherent peak in the nodal region and a broad hump 
in the antinodal region in underdoped Bi2212 samples below $T_{\rm c} $\cite{Tanaka}. From the quantitative agreement between theory and experiment we will conclude 
that the observed broad hump corresponds to the photo-electron excitations 
from the occupied states below the Fermi level to the free electron state above the vacuum level and thus the idea of a pseudogap in underdoped cuprates is no longer necessary. 

The third subject is the finite size effect of a metallic state in the K-S model on the spin-electronic structures of underdoped cuprates. In connection with the observed finite size of a spin-correlated region of the AF order \cite{Mason, Yamada}, Hamada and his coworkers \cite{Hamada} and Kamimura and Hamada \cite{KH} determined the ground state of the K-S model in a two-dimensional (2D) square lattice system with 16 (4 $\times$ 4) localized spins by the exact diagonalization method, and they clarified  that, in the presence of hole-carriers, 
the localized spins 
in a spin-correlated region tend to form an AF order rather than a random spin-singlet state and  thus that the hole-carriers can lower the kinetic energy of 
itineration in the spin-correlated region 
by taking the two kinds of orbitals alternately in the lattice of AF order. This mechanism of lowering the kinetic energy of the hole-carriers has led to the coexistence of a metallic state and the local AF order. This is the essential key-point of the K-S model. We call the above-mentioned behavior of a doped hole ``the mechanism of the K-S model''. In this paper we will suggest a possibility that a spatially inhomogeneous distribution of Fermi-pocket-states and of large-Fermi-surface states may appear due to the finite size effect when a temperature is higher than $T_{\rm c}$.  Finally a new phase diagram for underdoped cuprates is proposed.
 
  The organization of the present paper is the following: It consists of six parts. After Introduction, at the beginning of the Section II we will first describe the essential features of the K-S model which bears important characteristics born from the interplay of Jahn-Teller Physics and Mott Physics. Then we will show from the calculated many-body-effects included energy bands that the key features of the Fermi surfaces in underdoped cuprates are Fermi pockets. Further we will show that 
the ``Fermi arcs'' observed in ARPES is not a portion of a large Fermi surface, but it should be one of the edges of Fermi pockets in the nodal region. In Section III, on the basis of the many-body-effects included energy bands obtained from the K-S model we will predict the key features of ARPES spectra and clarify the origin of the two-gap scenario proposed from the experimental results by Tanaka {\it et al} \cite{Tanaka}. In Section IV we will discuss the finite size effects on the Fermi surfaces in cuprates, because a metallic state in the K-S model coexists with local antiferromagnetic (AF) order constructed from the localized spins whose spin-correlation length is finite. In connection with the finite size effects we will discuss a possibility of spatially inhomogeneous distribution of Fermi-pocket-states and of large-Fermi-surface states.  Taking account of the finite size effect, a new phase diagram is proposed in Section V. Section VI is devoted to conclusion and concluding remarks.

\section{Electronic structure of a metallic state in underdoped LSCO described by the K-S model}
In this section we will first describe the electronic structure of a metallic state in underdoped LSCO calculated by Kamimura and Suwa\cite{KS}, emphasizing the important roles due to the interplay of JahnTeller physics and Mott physics. 

Figure 1 shows the energy-level landscape starting from the orbitally doubly-degenerate e$_{\rm g}$ and triply-degenerate t$_{\rm 2g}$ states of a Cu$^{2+}$ ion in a CuO$_{6}$ octahedron with octahedral symmetry embedded in La$_2$CuO$_4$ at the left column. By the JT effect the Cu e$_{\rm g}$ orbital state splits into a$_{\rm 1g}$ and b$_{\rm 1g}$ orbital states, which form antibonding and bonding molecular orbitals of A$_{\rm 1g}$ and B$_{\rm 1g}$ symmetry with the molecular orbitals constructed from the in-plane oxygen p$_{\sigma}$ and apical oxygen p$_{z}$ orbitals in a CuO$_{6}$ octahedron with tetragonal symmetry, respectively. These molecular orbitals are denoted by a$^*_{\rm 1g}$, a$_{\rm 1g}$, b$_{\rm 1g}^{*}$ and b$_{\rm 1g}$, as shown at the middle column, where the asterisk {*} represents the antibonding orbital. In an undoped case, 7 electrons occupy these molecular orbitals, so that the highest occupied b$_{\rm 1g}^{*}$ state is half filled, resulting in an $S = 1/2$ state, where a b$_{\rm 1g}^{*}$ antibonding orbital has mainly Cu d$_{x^2-y^2}$ character. Following Mott physics, we introduce the Hubbard $U$ interaction ($U = 10$eV) as a strong electron-correlation effect. Then the half-filled b$_{\rm 1g}^{*}$ state splits into the lower and upper Hubbard bands denoted by L.H. and U.H. in the figure. The localized electrons in the L.H. band give rise to the localized spins around the Cu sites. These localized spins form the antiferromagnetic (AF) order by the superexchange interaction via intervening O$^{2-}$ ions in undoped La$_2$CuO$_4$. 

\begin{figure}[h]
\begin{center}
\includegraphics[width=9cm]{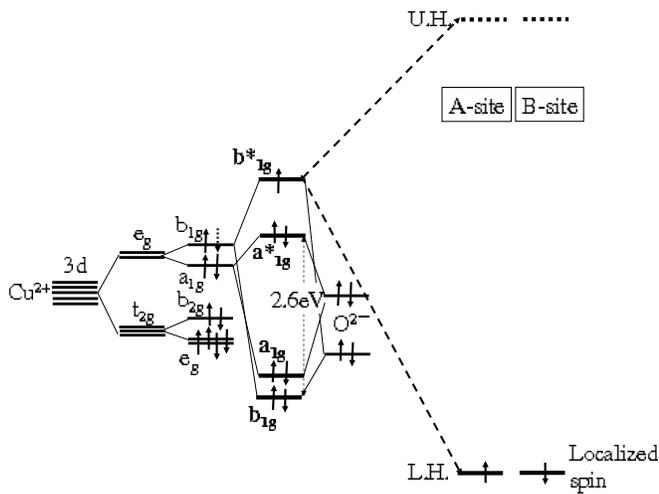}
\end{center}
\caption{\label{fig1}
Energy-level landscape due to the interplay of Jahn-teller physics and Mott physics, starting from the orbitally doubly-degenerate e$_{\rm g}$ and triply-degenerate t$_{\rm 2g}$ orbitals of a Cu$^{2+}$ ion in a regular CuO$_{6}$ octahedron with octahedral symmetry. The third column from the left shows the splitting of Cu e$_{\rm g}$ and t$_{\rm 2g}$ orbitals by the JT effect. The middle column shows the antibonding and bonding molecular orbitals of A$_{\rm 1g}$ and B$_{\rm 1g}$ symmetry formed by Cu a$_{\rm 1g}$ and b$_{\rm 1g}$ orbitals and the in-plane oxygen p$_{\sigma}$ and apical oxygen p$_{z}$ orbitals in an elongated CuO$_{6}$ octahedron with tetragonal symmetry. In the undoped case, these molecular orbitals accommodate 7 electrons. The right column shows the splitting of the highest half-filled b$_{\rm 1g}^{*}$ state into L.H. and U.H. bands by the Hubbard $U$ interaction. The spins of localized holes in the L.H. band at the neighboring A and B sites form an AF order. The energy in this figure is taken for electron-energy but not hole-energy}
\end{figure}

\subsection{Anti-Jahn-Teller Effect}
When Sr$^{2+}$ ions are substituted for La$^{3+}$ions in LSCO, one may think intuitively that apical oxygen (O$^{2-}$ ion) in the CuO$_{6}$ octahedrons tend to approach toward central Cu$^{2+}$ ions in order to gain the attractive electrostatic energy. Theoretically it was shown by the first-principles variational calculations of the spin-density-functional approach \cite{Shima, Oshiyama} that the optimized distance between apical O and Cu in La$_{2-x}$Sr$_x$CuO$_4$ (LSCO) which minimizes the total energy of LSCO decreases with increasing Sr concentration. As a result the elongated CuO$_{6}$ octahedrons by the Jahn-Teller (JT) interactions shrink by doping holes. This shrinking effect against the Jahn-Teller distortion is the ``anti-Jahn-Teller effect'' \cite{Copper Oxide}, as already explained in Introduction.

\begin{figure}[h]
\begin{center}
\includegraphics[width=9cm]{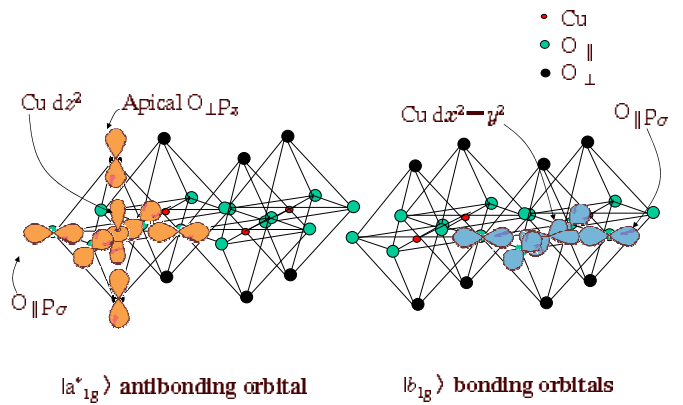}
\end{center}
\caption{\label{fig2}
Spatial extension of  $|$a$^*_{\rm 1g}$ $\rangle $ antibonding orbital and  $|$b$_{\rm 1g}$ $\rangle $ bonding orbital in a CuO$_{2}$ layer} 
\end{figure}

By this anti-Jahn-Teller effect, the energy separation between the two kinds of orbital states which have been split originally by the JT effect becomes smaller. These two states are the a$_{\rm 1g}$ anti-bonding orbital state $|{\rm a}_{\rm 1g}^*\rangle$ and the ${\rm b}_{\rm 1g}$ bonding orbital state $|{\rm b}_{\rm 1g}\rangle$ shown in the third column from the left in Fig.~\ref{fig1}. Here the 
a$^*_{\rm 1g}$ anti-bonding orbital state is constructed by Cu d$z^2$ orbital and six surrounding oxygen p orbitals including apical O p$_{z}$-orbitals while $|{\rm b}_{\rm 1g}\rangle$ orbital consists of four in-plane O p$_{\sigma}$ orbitals with a small Cu d$_{x^2-y^2}$ component parallel to a CuO$_2$ plane. The spatial extensions of  a$_{\rm 1g}^*$ and  ${\rm b}_{\rm 1g}$ orbitals are schematically shown in Fig.~\ref{fig2}.

\subsection{The important role of two kinds of many-electron states}
\begin{figure}[h]
\begin{center}
\includegraphics[width=9cm]{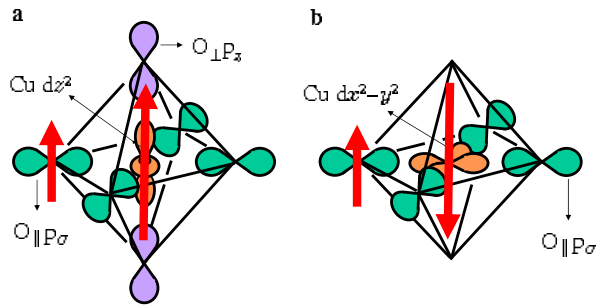}
\end{center}
\caption{\label{fig3}
Schematic pictures of Hund's coupling triplet $^3$B$_{\rm 1g}$ ({\bf a}) and Zhang-Rice singlet $^1$A$_{\rm 1g}$ ({\bf b}). The long and short arrows represent the spins of localized and doped holes, respectively.}
\end{figure}

 Kamimura and Eto calculated the lowest state energies of the many-electron states in which one hole is added into a CuO$_{6}$ octahedron embedded in LSCO \cite{KE, EK}. This means that an electron is taken out from the system shown in Fig.~\ref{fig1}. Hereafter we consider a hole as a doped carrier instead of the electron picture. In this case, there appear two kinds of many-particle states called multiplets. One case is that a dopant hole occupies an antibonding a$^*_{\rm 1g}$ orbital. In this case its spin becomes parallel to a localized spin in the b$_{\rm 1g}^{*}$ orbital to get a spin-triplet state, because the Hund's coupling makes the spins of holes in a$_{\rm 1g}^*$  and b$_{\rm 1g}^{*}$ orbitals with different symmetry parallel. As a result the energy of the a$^*_{\rm 1g}$ orbital state  for a single hole decreases by the Hund's coupling energy of 0.5 eV. Thus this spin-triplet multiplet is called the ``Hund's coupling triplet'' denoted by $^3$B$_{\rm 1g}$, which is schematically shown in Fig.~\ref{fig3}{\bf a}. 

The other case is that a dopant hole occupies a bonding b$_{\rm 1g}$ orbital, and its spin becomes antiparallel to the localized spin in the antibonding b$_{\rm 1g}^{*}$ orbital, since the dopant and localized holes occupy the orbitals of the same symmetry. As a result the energy of b$_{\rm 1g}$ orbital state for a single hole decreases by the spin-singlet exchange interaction of 3.0 eV. This spin-singlet multiplet may correspond to the ``Zhang-Rice singlet'' in the t-J model \cite{Zhang_Rice}, and it is denoted by $^1$A$_{\rm 1g}$. Zhang-Rice singlet state is schematically shown in Fig.~\ref{fig3}{\bf b}.

The energy separation between the two kinds of orbital states, the a$^*_{\rm 1g}$ anti-bonding orbital state and the b$_{\rm 1g}$ bonding orbital state, becomes smaller by the anti-Jahn-Teller effect when the hole concentration increases in the underdoped regime. Thus the Hund's coupling triplet state and the Zhang-Rice singlet state also appear at nearly the same energy in the underdoped region. 

By the first-principles cluster calculations which take into account the Madelung potential due to all the ions surrounding a CuO$_{6}$ cluster in LSCO and also the anti-Jahn-Teller effect, Kamimura and Eto showed that the lowest-state energies of these two multiplets are nearly equal when both the Hund's coupling exchange interaction and the spin-singlet exchange interaction are included \cite{KE,EK}. Consequently the energy difference between the highest occupied orbital states a$^*_{\rm 1g}$ in $^3$B$_{\rm 1g}$ multiplet and b$_{\rm 1g}$ in $^1$A$_{\rm 1g}$ multiplet becomes only 0.1eV for the optimum doping ($x = 0.15$) in La$_{2-x}$Sr$_{x}$CuO$_{4}$, as will be shown in the subsection D. 

\subsection{The explanation of the K-S model by the picture of a ``two-story house model''}
\begin{figure}
\begin{center}
\includegraphics[width=9cm]{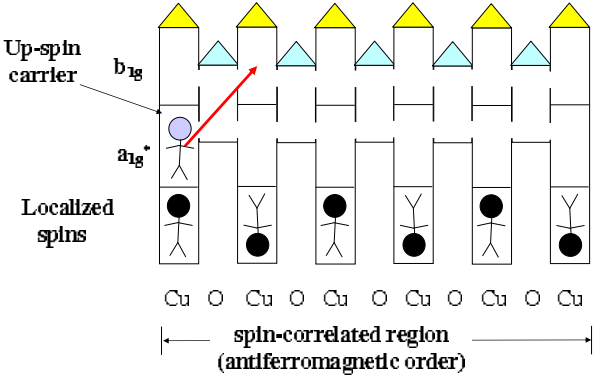}
\end{center}
\caption{\label{fig4}
Explanation of the K-S model by the picture of a two-story house model}
\end{figure}

By using the results of the first-principles cluster calculations for the lowest state of CuO$_{6}$ 
octahedron in LSCO by Kamimura and Eto mentioned in a previous subsection and assuming that the localized spins form an AF order by the superexchange interaction via intervening oxygen ions, Kamimura and Suwa \cite{KS} constructed a metallic state of LSCO for its underdoped regime. Since this model has already been explained in detail in refs. 6 and 7, here we explain the key-features of the model in a heuristic way using the picture of a two-story house model shown in Fig.\ref{fig4}. In this figure the first story of a Cu house with (yellow) roof is occupied by the Cu localized spins, which form the AF order in the spin-correlated region by the superexchange interaction. 

The second story in a Cu house consists of two floors due to the anti-JT effect, lower a$^*_{\rm 1g}$ floor and upper b$_{\rm 1g}$ floor. The second story between neighboring Cu houses are connected by oxygen rooms with (blue-color) roof, reflecting the hybridization of Cu d and O p orbitals. In the second story a hole-carrier with up spin enters into the a$^*_{\rm 1g}$ floor at the left-hand Cu house due to Hund's coupling with Cu localized up-spin in the first story (Hund's coupling triplet), as shown in the extreme left column of the figure. By the transfer interaction marked by a long (red-color) arrow in the figure, the hole is transferred into the b$_{\rm 1g}$ floor at the neighboring Cu house (the second from the left) through the oxygen room, where the hole with up spin forms a spin-singlet state with a localized down spin at the second Cu house from the left (Zhang-Rice singlet). The key feature of the K-S model is that the hole-carriers in the underdoped regime of LSCO form a metallic state, by taking the Hund's coupling triplet and the Zhang-Rice singlet alternately in the presence of the local AF order without destroying the AF order, as shown in the figure. Since the second story consists of the two floors of different symmetry, the  two-story house model represents the two-component 
theory. As seen in Fig.\ref{fig4}, the characteristic feature of the K-S model is the coexistence of the AF order and a normal, metallic (or a superconducting) state in the underdoped regime. This feature in the K-S model (two-component theory) is different from that of the single-component theory.

\subsection{Effective Hamiltonian for the K-S model}
The following effective Hamiltonian is introduced in order to describe the ``K-S model'' following Kamimura and Suwa \cite{KS} (see also ref.~\cite{Kamimura_Entropy}). 
It consists of four parts: The one-electron Hamiltonian $H_{\rm sing}$ 
for a$_{\rm 1g}^{\ast}$ and b$_{\rm 1g}$ orbital states, 
the transfer interaction between neighboring CuO$_6$ octahedrons $H_{\rm tr}$, 
the superexchange interaction between the Cu $d_{x^2-y^2}$ localized spins $H_{\rm AF}$, 
and the exchange interactions between the spins of dopant holes and of $d_{x^2-y^2}$ localized holes 
within the same CuO$_6$ octahedron $H_{\rm ex}$.
Thus we have 
\begin{eqnarray}
H &=& H_{\rm sing} + H_{\rm tr} + H_{\rm AF} + H_{\rm  ex}  \nonumber \\
  &=& \sum_{i,m,\sigma} \varepsilon_m C_{im\sigma}^\dagger C_{im\sigma} 
\nonumber \\
    & & {}+ \sum_{\langle i,j\rangle,m,n,\sigma} t_{mn}
              \left(C_{im\sigma}^\dagger C_{jn\sigma} + {\rm h.c.} \right) \nonumber \\ 
  & & {} + J \sum_{\langle i,j\rangle} {\bm{S}}_i \cdot {\bm{S}}_j 
         + \sum_{i,m} K_m\, {\bm{s}}_{i,m} \cdot {\bm{S}}_i \ \ , \label{eq:H}
\end{eqnarray}
where $\varepsilon_m$ ($m=$ a$_{\rm 1g}^{\ast}$ or b$_{\rm 1g}$) 
represents the one-electron energy of 
the a$_{\rm 1g}^{\ast}$ and b$_{\rm 1g}$ orbital states, 
$C_{im\sigma}^\dagger$ and $C_{im\sigma}$ are the creation and annihilation operators of a dopant hole 
with spin $\sigma$ in the $i$-th CuO$_6$ octahedron, respectively, 
$t_{mn}$ the transfer integrals of a dopant hole between $m$-type and $n$-type orbitals 
of neighboring CuO$_6$ octahedrons, $J$ the superexchange interaction 
between the spins ${\bm{S}}_i$ and ${\bm{S}}_j$ of $d_{x^2-y^2}$ localized holes 
in the b$_{\rm 1g}^{\ast}$ orbital at the nearest neighbor Cu sites $i$ and $j$ 
($J > 0$ for AF interaction), 
and $K_m$ the exchange integrals for the exchange interactions 
between the spin of a dopant hole ${\bm{s}}_{im}$ and the $d_{x^2-y^2}$ localized spin ${\bm{S}}_i$ 
in the $i$-th CuO$_6$ octahedron. There are two exchange constants, $K_{\rm a_{\rm 1g}^{\ast}}$  and
 $K_{\rm b_{\rm 1g}}$, for the Hund's coupling triplet and Zhang-Rice singlet, respectively, where 
 $K_{\rm a_{\rm 1g}^{\ast}} < 0$ and $K_{\rm b_{\rm 1g}} > 0$. The appearance of the two kinds of 
 exchange interactions in the fourth term is the key-feature of the K-S model.

By replacing the localized spins ${\bm{S}}_i$'s in $H_{\rm ex}$ by their average value $\langle {\bm{S}}\rangle$ in the mean-field sense, we can calculate the change of the total energy upon moving of a hole from an a$^*_{\rm 1g}$ orbital state in the Hund's coupling spin-triplet at Cu site $i$ to an empty b$_{\rm 1g}$ orbital state in the Zhang-Rice spin-singlet at the neighbouring Cu site $j$. At the first step the hole moves from Cu site $i$ to infinity. The change of the total energy in the mean field approximation is equal to $\varepsilon_{{\rm a}_{\rm 1g}^{\ast}}+ \frac{1}{4}K_{{\rm a}_{\rm 1g}^{\ast}}$. At the second step the hole moves from infinity to an empty b$_{\rm 1g}$ orbital state at Cu site $j$ to form the Zhang-Rice singlet. The change of the total energy in the second step is equal to $\varepsilon_{\rm b_{\rm 1g}} - \frac{3}{4}K_{\rm b_{\rm 1g}}$. As a result the change of the total energy by the transfer of the hole from the occupied ${\rm a_{\rm 1g}^{\ast}}$ orbital state at Cu site $i$ to the empty ${\rm b_{\rm 1g}}$ orbital state at Cu site $j$ is 
\begin{eqnarray}
  \varepsilon_{\rm a_{\rm 1g}^{\ast}}^{\rm ~eff} -  \varepsilon_{\rm b_{\rm 1g}}^{\rm ~eff} = \varepsilon_{\rm a_{\rm 1g}^{\ast}} + \frac{1}{4}K_{\rm a_{\rm 1g}^{\ast}} - \varepsilon_{\rm b_{\rm 1g}} + \frac{3}{4}K_{\rm b_{\rm 1g}}, \label{eq:D}.        
\end{eqnarray}

Here $\varepsilon_{\rm a_{\rm 1g}^{\ast}}^{\rm ~eff}$ and $\varepsilon_{\rm b_{\rm 1g}}^{\rm ~eff}$ represents the effective one electron energy of a$_{\rm 1g}^{\ast}$ and b$_{\rm 1g}$ orbital states including the exchange interaction term $H_{\rm  ex}$, respectively. Thus the energy difference $(\varepsilon_{\rm a_{\rm 1g}^{\ast}}^{\rm ~eff} - \varepsilon_{\rm b_{\rm 1g}}^{\rm ~eff})$ corresponds to
the energy difference between ${\rm a_{\rm 1g}^{\ast}}$ and ${\rm b_{\rm 1g}}$ floors in the second story in Fig.~\ref{fig4}.

\begin{figure}[h]
\begin{center}
\includegraphics[width=9cm]{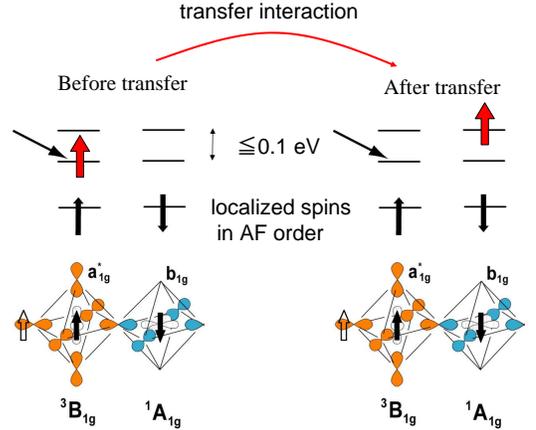}
\end{center}
\caption{\label{fig5}
Energy diagrams of a$^*_{\rm 1g}$ orbital state and  the b$_{\rm 1g}$ orbital state at the neighbouring Cu sites $i$ and $j$ before and after the transfer of a hole from Cu site $i$ to $j$, where the localized spins at $i$ and $j$ sites are also shown. Energy in this figure is taken for the hole-energy.}
\end{figure}

Now let us estimate the energy difference, $\varepsilon_{\rm a_{\rm 1g}^{\ast}}^{\rm ~eff}$ -  $\varepsilon_{\rm b_{\rm 1g}}^{\rm ~eff}$, by using the values of the parameters in the effective Hamiltonian~(\ref{eq:H}). The values of the parameters in equation~(\ref{eq:H}) have been determined in the case of LSCO in ref.~\cite{KS} (see also ref.~\cite{Kamimura_Entropy}). Those are: $J=0.1$, 
$K_{\rm a_{\rm 1g}^{\ast}}=-2.0$, $K_{\rm b_{\rm 1g}}=4.0$, 
$t_{\rm a_{\rm 1g}^{\ast} a_{\rm 1g}^{\ast}}=0.2$, 
$t_{\rm b_{\rm 1g} b_{\rm 1g}}=0.4$, 
$t_{\rm a_{\rm 1g}^{\ast} b_{\rm 1g}} 
= \sqrt{t_{\rm a_{\rm 1g}^{\ast} a_{\rm 1g}^{\ast}} t_{\rm b_{\rm 1g} b_{\rm 1g}}} \sim 0.28$, 
$\varepsilon_{\rm a_{\rm 1g}^{\ast}}=0$, $\varepsilon_{\rm b_{\rm 1g}}=2.6$ in units of eV, 
where the values of $K_{\rm a_{\rm 1g}^{\ast}}$ and $K_{\rm b_{\rm 1g}}$ are taken from the first principles cluster calculations 
for a CuO$_{6}$ octahedron in LSCO \cite{KE, EK}, and the values of $t_{mn}$'s are obtained from band structure calculations \cite{Shima, Oshiyama}. 
The energy difference of the one-electron energies 
between a$_{\rm 1g}^{\ast}$ and b$_{\rm 1g}$ orbital states in a CuO$_{6}$ octahedron for a certain value of $x$ has been determined so as to reproduce 
the difference of the lowest state energies between 
the Hund's coupling spin-triplet state and the Zhang-Rice spin-singlet state for the same value of $x$ in LSCO calculated by the MCSCF cluster calculations which includes the anti-JT-effect \cite{KS}. 

Thus the calculated value of $\varepsilon_{\rm a_{\rm 1g}^{\ast}}^{\rm ~eff}$ -  $\varepsilon_{\rm b_{\rm 1g}}^{\rm ~eff}$ is 0.1 eV for the case of the optimum doping ($x = 0.15$). Then, by introducing the transfer interaction of $t_{\rm a_{\rm 1g}^{\ast} b_{\rm 1g}} = 0.28$, a coherent metallic state in the normal phase is obtained under the coexistence with the local AF order for the underdoped regime. This situation is schematically shown in Fig.~\ref{fig5}, According to Kamimura and Eto \cite{KE, EK}, 
the ground state energy of the Hund's coupling spin-triplet is nearly equal to that of the Zhang-Rice spin-singlet state for any value of $x$ in the underdoped regime, so that the energy difference between the a$^*_{\rm 1g}$ orbital state and  the b$_{\rm 1g}$ orbital state at the neighbouring Cu sites $i$ and $j$, $\varepsilon_{\rm a_{\rm 1g}^{\ast}}^{\rm ~eff}$ -  $\varepsilon_{\rm b_{\rm 1g}}^{\rm ~eff}$, is almost zero in the underdoped regime.

\subsection{Features of the many-body-effects included energy bands and Fermi surfaces of underdoped LSCO coexisting with the AF order}
In a previous subsection we have shown that the effective Hamiltonian (\ref{eq:H}) for the K-S model can lead to a unique metallic state in the normal phase which results in of the coexistence of a superconducting state and AF order below $T_{\rm c}$. In 1994 Kamimura and Ushio calculated the energy bands and Fermi surfaces of underdoped LSCO in the normal phase, based on the effective Hamiltonian (\ref{eq:H}), by treating the fourth term $H_{\rm ex}$ in the effective Hamiltonian (\ref{eq:H}) by the mean-field approximation, that is by replacing the localized spins ${\bm{S}}_i$'s by their average value $\langle {\bm{S}}\rangle$ \cite{KU, UK}. Thus the effect of the localized spin system was dealt with as an effective magnetic field acting on the hole carriers. As a result they could separate the localized hole-spin system in the AF order and the hole-carrier system from each other, and calculated the ``one-electron type'' energy band for a carrier system assuming a periodic AF order. Here ``one-electron type'' is meant by the inclusion of many-body-effects in the energy bands. That is, the exchange interactions between carrier's and localized spins are included by the mean field approximation.

\begin{figure}
\begin{center}
\includegraphics[width=9cm]{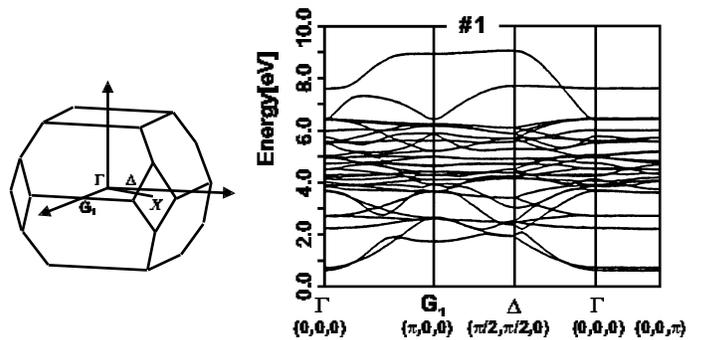}
\caption{\label{fig6}The many-body-effect included band-structure for up-spin (or down spin) dopant holes. The highest occupied band is marked by the \#1 band (right) and the AF-Brillouin zone (left). The $\Delta$-point corresponds to $(\pi /2a, \pi /2a,0)$, 
while the G$_1$-point corresponds to $(\pi /a,0,0)$. 
}
\end{center}
\end{figure}

In Fig.~\ref{fig6}, the calculated many-body-effect included energy band structure for up-spin (or down-spin) doped holes for LSCO is shown for various values of wave-vector $\bm{k}$ and symmetry points in the antiferromagnetic (AF) Brillouin zone, where the AF Brillouin zone is adopted because of the coexistence of a metallic state and the AF order, and it is shown at the left side of the figure. Here one should note that the energy in this figure is taken for electron-energy but not hole-energy. Further the Hubbard bands for localized b$^{\ast}_{\rm 1g}$  holes which contribute to the local AF order are separated and do not appear in this figure. 

In the undoped La$_{2}$CuO$_{4}$, all the energy bands in Fig.~\ref{fig6} are occupied fully by electrons so that La$_{2}$CuO$_{4}$ is an antiferromagnetic Mott insulator, consistent with experimental results. In this respect the present effective energy band structure is completely different from the ordinary LDA energy bands \cite{Mattheiss,Yu}. When Sr are doped, holes begin to occupy the top of the highest band in Fig.~\ref{fig6} marked by \#1 at $\Delta$ point which corresponds to $(\pi /2a, \pi /2a, 0)$ in the AF Brillouin zone. At the onset concentration of superconductivity, the Fermi level is located just below the top of the \#1 band at $\Delta$, which is a little higher than that of the G$_1$ point. Here the G$_1$ point in the AF Brillouin zone lies  at $(\pi /a, 0, 0)$, and corresponds to a saddle point of the van Hove singularity.

\begin{figure}
\includegraphics[width=4cm]{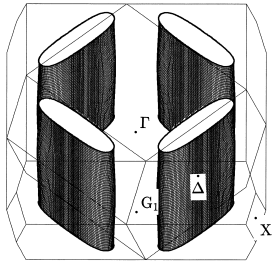}
\caption{\label{fig7}
The Fermi surface of hole-carriers
 for $x=0.15$ calculated for the \#1 band. 
Here two kinds of Brillouin zones are also shown. 
One at the outermost part is the ordinary Brillouin zone 
corresponding to an ordinary unit cell consisting of a single CuO$_6$ octahedron and the inner part is the Brillouin zone for the AF unit cell in LSCO. 
Here the $k_x$ axis is taken along $\overline{\Gamma {\rm G}_1}$, 
corresponding to the $x$-axis (the Cu--O--Cu direction) in a real space.}
\end{figure}

Based on the calculated band structure shown in Fig.~\ref{fig6}, Kamimura and Ushio  \cite{KU, UK} calculated the Fermi surfaces 
for the underdoped regime of LSCO. In Fig.~\ref{fig7} the Fermi surface structure of hole-carriers calculated for $x = 0.15$ is shown as an example, where the Fermi surface (FS) consists of four Fermi pockets of extremely flat tubes around $\Delta$ point, $(\pi /2a, \pi /2a, 0)$, and the other three equivalent points (the nodal region) in the momentum space. The total volume of the four Fermi pockets is proportional to the concentration of the doped hole-carriers. Thus the feature of Fermi pockets constructed from the doped holes shown in Fig.~\ref{fig7} is consistent with Luttinger's theorem in the presence of AF order \cite{Luttinger}.

In 1996 to 1997 Mason {\it et al} \cite{Mason} and Yamada {\it et al} \cite{Yamada} reported independently the magnetic coherence effects on the metallic and superconducting states in underdoped LSCO by neutron inelastic scattering measurements.  Since then a number of papers suggesting the coexistence of local AF order and superconductivity in cuprates by neutron and NMR experiments have been published \cite{Yamada_1998, Kao, Christensen, Tranquada, Haydon, Mukuda}. 

The Fermi surface structure in Fig.~\ref{fig7} is completely different from that of the single components theory, in which a Fermi surface is large. Recently Meng and his coworkers reported the existence of the Fermi pocket structure in the ARPES measurements of underdoped Bi$_2$Sr$_{2-x}$La$_x$CuO$_{6+\delta}$ (La-Bi2201) \cite{Meng}. Their results are the clear experimental evidence for our prediction of the Fermi-pocket-structure for underdoped LSCO in 1994 \cite{KU}.

It is interesting to see a many-body effect on the Fermi-pockets as follows: By averaging the character of wavefunction for each wave-vector $\bm{k}$ 
over the Fermi surface for $x=0.1$ in LSCO, 
we have obtained the mixing ratio of characters of the Hund's coupling triplet state 
to the Zhang-Rice singlet state in the metallic state at the Fermi level, 
which is found to be 1 to 5. 
Thus the existence ratio of the Hund's coupling triplet state in the metallic 
state at the Fermi level is only 20 \%. 
This value is close to the observed intensity ratio of the $\bm{E} \parallel c$ 
to the $\bm{E} \perp c$ polarization of x-ray at the Cu L$^{\prime}_{3}$ satellite peak 
for $x=0.1$ in LSCO in the polarized x-ray absorption spectra reported by Chen {\it et al} \cite{Chen}. 

In 1997 Anisimov, Ezhov and Rice calculated the energy band structure 
of the ordered alloy La$_{2}$Li$_{0.5}$Cu$_{0.5}$O$_{4}$ by the LDA+U method \cite{Anisimov}, 
and they showed that a fairly modest reduction of the apical Cu-O bond length is sufficient to stabilize 
the Hund's coupling spin triplet state with dopant holes in both b$_{\rm 1g}$ and a$^{\ast}_{\rm 1g}$ orbitals. Their calculated result supports the K-S model.

\section{The characteristic features of ARPES spectra obtained from the K-S 
model and the conclusion of the absence of pseudogap}
Recently considerable attention has been paid to the phenomenological idea of a pseudogap. When a portion of the Fermi surface in cuprates were not seen in the ARPES experiments, an idea of pseudogap was proposed as a kind of gap to truncate the Fermi surface in the single particle spectrum \cite{Marshall, Norman_Nature}. The disconnected segments of the Fermi surface are called ``Fermi arc'' \cite{Norman_Nature, Yoshida1, Norman_PRB2007}. Further ARPES experiments reported that such pseudogap develops below a temperature called $T^*$ which depends on the hole concentration $x$ in the underdoped regime of cuprates, and thus we write $T^*(x)$ hereafter. The $T^*(x)$ decreases with increasing the hole concentration $x$ and disappears at a certain concentration $x_{\rm o}$ in the overdoped region \cite{Kanigel}. In this section, on the basis of the K-S model we clarify the origins of the pseudogap and of $T^*(x)$.

\subsection{Features of the calculated ARPES spectra, clarification of 
the observed two-gap scenario, and the absence of pseudogap}

\begin{figure}
\begin{center}
\includegraphics[width=9cm]{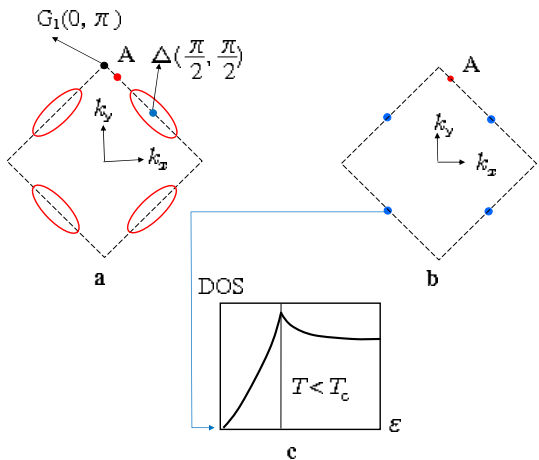}
\end{center}
\caption{
\label{fig9}
The 2D projected Fermi pockets in the nodal region above $T_{\rm c}$ ({\bf a}), their change to the d-wave nodes below $T_{\rm c}$ ({\bf b}) and the d-wave superconducting density of states ({\bf c}.)}
\end{figure}

Following the K-S model, the Fermi surface in the underdoped regime is the Fermi pockets in the nodal region as shown in Fig.~\ref{fig7}. Here we project the 3D picture of the Fermi pockets in Fig.~\ref{fig7} on the $k_{x}$-$k_{y}$ plane in the momentum space. The projected 2D picture of the four Fermi pockets around $\Delta$ point, $(\pi /2a, \pi /2a)$, and the other three equivalent points in the momentum space is shown on the antiferromagnetic Brillouin zone in Fig.~\ref{fig9}({\bf a}). Below $T_{\rm c}$ the hole carriers in these Fermi pockets form Cooper pairs, contributing to the formation of a superconducting state, and a superconducting gap appears across the Fermi level. This feature is consistent with the Uemura plot \cite{Uemura}. In Fig.~\ref{fig9}({\bf b}) the d-wave node below $T_{\rm c}$ predicted by the K-S model \cite{Kamimura_d-Wave, Kamimura_Super} is schematically shown as dots, and the d-wave superconducting density of states (DOS) is schematically shown in Fig.~\ref{fig9}({\bf c}). Here one should note that AF order still coexists with a superconducting state below $T_{\rm c}$ so that we can use the same antiferromagnetic Brillouin zone, as seen in Fig.~\ref{fig9}({\bf b}).

\begin{figure}
\begin{center}
\includegraphics[width=9cm]{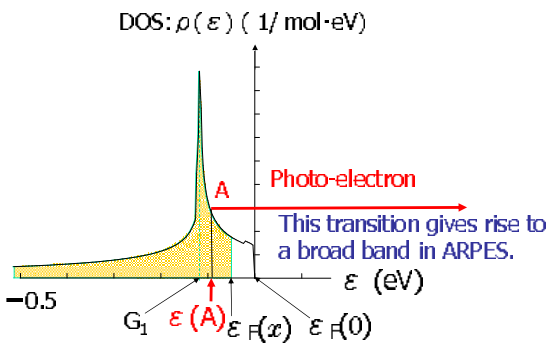}
\end{center}
\caption{
\label{fig10}
Schematic picture illustrating the real transitions of photo-excited electrons from state A below the Fermi level $\varepsilon _{\rm F}$ to a free-electron state above the vacuum level. This is the origin of a broad hump in the ARPES spectra, reflecting the density of occupied states shaded in this figure. The broad hump appears in the antinodal region around the G$_1$-point in the momentum space.}
\end{figure}

On the other hand, in the antinodal region the states occupied by electrons which do not participate in the formation of superconductivity still exist below $T_{\rm c}$. As an example of such states, a state A is exemplified in Fig.~\ref{fig9}({\bf b}), and the state corresponding to A above $T_{\rm c}$ is also shown in Fig.~\ref{fig9}({\bf a}).  Then the real transitions of electrons from the occupied states, say the state A, below the Fermi level $\varepsilon _{\rm F}$ in the \#1 energy band in Fig.~\ref{fig6} to a free-electron state above the vacuum level occur by photo-excitation both above and below $T_{\rm c}$ around the G$_1$-point, $(\pi /a,0,0)$, and other equivalent points in the momentum space. These transitions appear in the antinodal region in the momentum space. Such transitions are illustrated in Fig.~\ref{fig10}, where the density of states (DOS) $\rho $($\varepsilon $) calculated for the \#1 energy band by Ushio and Kamimura \cite{UK} is shown.  Since the intensity of such photo-excited transitions is proportional to the density of occupied states below $\varepsilon _{\rm F}$, we can predict that such photo-excited transitions appear as a broad hump with a peak in ARPES spectra, reflecting the density of occupied states below $\varepsilon _{\rm F}$ shaded in Fig.~\ref{fig10}.

Based on the situation shown in Fig.~\ref{fig9}({\bf a}) to Fig.~\ref{fig9}({\bf c}) and Fig.~\ref{fig10}, we can predict the features of ARPES spectra below $T_{\rm c}$ theoretically in the following way: A coherent peak due to the superconducting density of states appears at the nodal region around $\Delta$ point while a broad hump with a peak reflecting the shaded area of DOS in Fig.~\ref{fig10} 
appears in a wide region centered at the antinodal point G$_1$ in the momentum space. The latter corresponds to the real transitions of electrons from the occupied states below the Fermi level $\varepsilon _{\rm F}$ to a free-electron state above the vacuum level, with the intensity proportional to the density of states (DOS) $\rho $($\varepsilon $) below the Fermi level $\varepsilon _{\rm F}$ shown in Fig.~\ref{fig10}, where $\varepsilon _{\rm F}$  varies with the hole-concentration $x$, so that we write $\varepsilon _{\rm F}(x)$ hereafter. Thus we designate a broad hump as ``antinodal transitions''. This theoretical ARPES spectra are very similar to 
those reported by Tanaka {\it et al}. for Bi2212 \cite{Tanaka}, where the experimental results revealed two distinct energy gaps exhibiting different doping dependence.

\begin{figure}[h]
\begin{center}
\includegraphics[width=9cm]{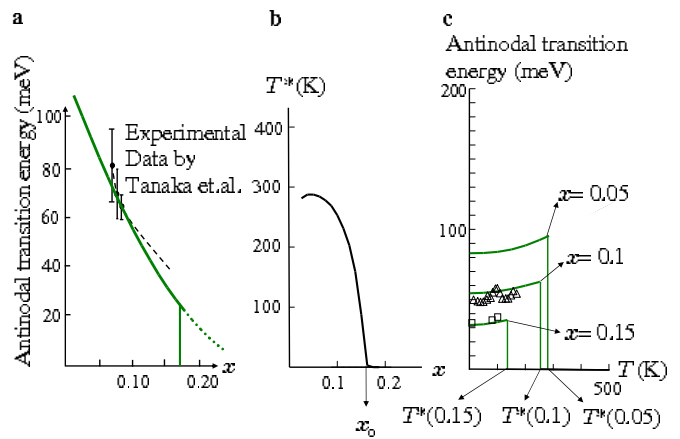}
\end{center}
\caption{\label{fig11}
Calculated ARPES spectra for LSCO and comparison with experimental results of Bi2212. \textbf{a,} Calculated energy difference $|\varepsilon ({\rm G}_1)-\varepsilon _{\rm F}(x)|$ as a function of the hole concentration $x$ at $T = 0$K. Experimental results by Tanaka {\it et al} \cite{Tanaka} are shown by dots. \textbf{b,} Calculated $T^* (x)$ is shown by a solid line. \textbf{c,} Calculated temperature dependence of the antinodal transition energy for $x$ = 0.05, 0.1, and 0.15. Experimental data by Norman {\it et al} \cite{Norman_Phys_Rev} (triangle) and Lee {\it et al} \cite{Lee} (square) are shown.}
\end{figure}

From ARPES experiments for the ``antinodal transitions'' one can measure the energy difference between the energy $\varepsilon({\rm A})$ of state A and the Fermi energy $\varepsilon_{\rm F}$, $|\varepsilon({\rm A})-\varepsilon_{\rm F}(x)|$, shown in Fig.~\ref{fig10}. Since the ${\rm G}_1$ point at the edge of the antiferromagnetic BZ corresponds to a saddle-point singularity, we may expect the appearance of a peak in the broad hump for the transition from the  ${\rm G}_1$ point. In this context we have calculated the doping dependence of the energy difference  $|\varepsilon({\rm G}_1)-\varepsilon_{\rm F}(x)|$ for LSCO. Here we designate $|\varepsilon({\rm G}_1)-\varepsilon_{\rm F}(x)|$ as ``the antinodal transition energy''. The Fermi energy in the undoped case, $\varepsilon _{\rm F}(0)\equiv\varepsilon _{\rm F}(x=0)$,  lies at the right edge of $\rho (\varepsilon )$ (DOS) in Fig.~\ref{fig10}, and thus the antinodal transition energy decreases with increasing $x$. Since a shape of DOS for the highest conduction band does not depend on cuprate materials, we compare in Fig.~\ref{fig11}{\bf a} the calculated doping dependence of the antinodal transition energy for LSCO (solid lines) with the experimental results of ``antinodal gap'' for Bi2212 in ref.~\cite{Tanaka} which are shown as dots in the figure. As seen in Fig.~\ref{fig11}{\bf a}, the agreement between theory and experiment is remarkably good. Here we would like to emphasize that there are no disposal parameters in the present case. From this quantitative agreement we can conclude that among the observed two gaps below $T_{\rm c}$, a gap associated with antinodal regime corresponds to the real transitions of electrons from the occupied states below the Fermi level to a free-electron state above the vacuum level while the other gap associated with the near-nodal regime corresponds to the superconducting gap created on the Fermi pockets. From the marvelous agreement between the present theory and experimental results by Tanaka and his coworkers, we can conclude that the pseudogap is absent in cuprates.

Recently Yang {\it et al} \cite{Yang} have suggested from their ARPES experiments on Bi2212 that the opening of a symmetric gap related to superconductivity occurs only in the antinodal region and that the pseudogap reflects the formation of preformed pairs, contrary to the ARPES experimental results reported by Tanaka {\it et al} \cite{Tanaka}. In the present theory we have clearly shown in Fig.~\ref{fig9} and Fig.~\ref{fig11} that, in the ARPES experiments a peak related to superconductivity appears only in the nodal region and the spectra in the antinodal region correspond to the photoexcitations from the occupied states below $\varepsilon_{\rm F}$ to a free-electron state above the vacuum. If the antinodal region in ref.~\cite{Yang} means the region around the ${\rm G}_1$ point in the present paper, their suggestion is different from our theoretical results. Finally we should remark that any proposed theory must explain both the doping and temperature dependences of ARPES spectra in the underdoped regime consistently. From this standpoint we will investigate the temperature dependence of the ARPES spectra theoretically in the next subsection.

\subsection{Physical meaning of $T^* (x)$ and the temperature 
dependence of ARPES spectra}

For the purpose of calculating the temperature dependence of the ``antinodal transition energy'', first of all we would like to clarify the physical meaning of $T^* (x)$. When a hole concentration $x$ is fixed at a certain value in the underdoped region and temperature increases beyond $T_{\rm c}$, at a certain temperature in the normal phase the local AF order constructed by superexchange interaction in a CuO$_2$ plane becomes destroyed thermally and thus the coexistence of a metallic state with the Fermi pockets and of the local AF order will disappear. As a result the K-S model will not hold at that temperature in the underdoped regime. This temperature is defined as $T^* (x)$. Thus the phase of the Fermi pockets coexisting with the local AF order in the K-S model holds only below $T^* (x)$. We designate the phase of the Fermi pockets as the ``small Fermi-surface (FS)'' phase. Hereafter it is abbreviated as the SF-phase. When temperature increases beyond $T^* (x)$, a system changes from the SF-phase to the electronic phase consisting of a large FS without the AF order. We call this phase the large FS-phase, abbreviated as the LF-phase. 

In this context we may consider that the $T^* (x)$ represents the boundary between the SF- and LF- phases. Now let us introduce a quantity defining the difference between the free energies of the LF- and SF- phases:
\begin{eqnarray}
\Delta F(T, x)&\equiv &F_{\rm LF}(T, x)-F_{\rm SF}(T, x),
\label{eq1}
\end{eqnarray}
where $F_{\rm LF}(T, x)$ and $F_{\rm SF}(T, x)$ are the free energies of LF- and SF- phases, respectively. Here the free energy $F(T, x)$  is defined as 
\begin{eqnarray}
F(T, x)&=&E (T, x)-TS(T, x),
\label{eq2}
\end{eqnarray}
where $E(T, x)$ and $S(T, x)$ are the internal energy and entropy of each phase, respectively. These quantities are calculated from 
\begin{eqnarray}
 E(T, x) &=& \int ^{\infty}_{-\infty}\varepsilon \rho (\varepsilon ) f(\varepsilon , \mu(x)) {\rm d}\varepsilon , 
\label{eq3a}
\end{eqnarray}
and
\begin{eqnarray}
S(T,x) &=& - k_{\rm B} \int^{\infty}_{-\infty} \Bigl[ f(\varepsilon,\mu(x)) \ln f(\varepsilon,\mu(x)) \nonumber \\
       & & {}+ \bigl\{ 1-f(\varepsilon,\mu(x)) \bigr\} \ln \bigl\{ 
1-f(\varepsilon,\mu(x)) \bigr\} \Bigl]~
\nonumber \\
        & & ~~~~~~~~~~~~~~~~~~~~~~~~~~~~~~~~~~~~~~ \rho(\varepsilon)~ d\varepsilon \label{eq:S} \,,
\label{eq3b}
\end{eqnarray}
where $\mu (x)$ is a chemical potential of each phase and $f(\varepsilon , \mu(x))$ the Fermi distribution function at energy $\varepsilon$ and chemical potential $\mu (x)$ . Then $T^* (x)$ is defined by the equation 
\begin{eqnarray}
\Delta F(T^* (x), x) = 0. 
\label{eq4}
\end{eqnarray}
Kamimura, Hamada and Ushio calculated the electronic entropies for the SF- and LF- phases of LSCO \cite{Kamimura_Entropy}. According to their results the difference of electronic entropy between the SF- and LF- phases increases with increasing the hole-concentration $x$ in the underdoped regime. Using this result we find that 
$T^* (x)$ calculated from equation (\ref{eq4}) decreases with increasing $x$, as shown in Fig.~\ref{fig11}{\bf b}. In this calculation we have introduced two parameters $T^*(x=0.05)$ and $x_{\rm o}$, where $x_{\rm o}$ 
means the critical concentration that satisfies $T^*(x_{\rm o})$=0, and $T^*(x=0.05)$ represents a quantity related to the energy difference between the phase of doped AF insulator and the LF phase at the onset concentration of the metal-insulator transition ($x=0.05$). In Fig.~\ref{fig11}{\bf b} we have chosen $T^*(x=0.05)$ to be 300K \cite{Miyakawa}. 

On the other hand, the physical meaning of $T^*(x_{\rm o})=0$ can be explained in the following way: When the hole concentration exceeds the optimum doping ($x$ = 0.15) for LSCO and enters a slightly overdoped region, the local AF order via superexchange interaction in a CuO$_2$ plane begins to be destroyed by an excess of hole-carriers. Thus the K-S model does not hold at a certain concentration $x_{\rm o}$ in the overdoped region, and hence the small FS in the K-S model changes to a large FS. Thus $T^* (x)$ vanishes at $x_{\rm o}$. From the analysis of various experimental results we choose $x_{\rm o}$ = 0.17 for LSCO \cite{Nakano, Cooper}.  From this explanation we can say that the area below $T^* (x)$ in the underdoped regime represents the region in which the normal, metallic phase above $T_{\rm c}$ and the superconducting phase below $T_{\rm c}$ coexist with the local AF order.

In this circumstance it is clear that the ``antinodal transition energy'' defined by  $|\varepsilon({\rm G}_1)-\varepsilon_{\rm F}(x) |$  appears at temperatures below $T^*(x)$ and vanishes at $T^*(x)$. Now we calculate the temperature dependence of the antinodal transition energy by using equations (\ref{eq1}), (\ref{eq2}) , (\ref{eq3a}) and (\ref{eq3b}). The calculated results for three concentrations $x$ = 0.05, 0.10 and 0.15 of LSCO in the underdoped to optimum-doped region are shown in Fig.~\ref{fig11}{\bf c} as a function of temperature, where the parameter $T^*(0.05)$ = 300K is used.  As seen in the figure, the antinodal transition energy increases slightly with temperature up to $T^*(x)$ and vanished sharply at $T^*(x)$. These calculated results are compared with experimental results of underdoped sample of Bi2212 in refs.~\cite{Norman_Phys_Rev} and \cite{Lee}, which are indicated by triangle and square, respectively, in Fig.~\ref{fig11}{\bf c}. As seen in the figure, the agreement between theory and experiment is remarkably good. 

 \section{Spatially inhomogeneous distribution of Fermi-pocket-states and of 
 large-Fermi-surface-states due to the finite size effects}
\subsection{Finite size effects of a metallic state on Fermi surfaces}
According to the results of neutron inelastic scattering experiments by 
Mason \cite{Mason} and Yamada \cite{Yamada}, the AF spin-correlation length $\lambda_{s}$ in the underdoped region of LSCO is finite. In the underdoped regime of LSCO it increases as the Sr concentration increases from $x = 0.05$ in LSCO, 
the onset of superconductivity, and reaches a value of about 50\AA ~or more at the optimum doping ($x = 0.15$). In this subsection we discuss the effects of 
the finite size of the AF spin-correlation length on the structure of Fermi pockets shown in Fig.~\ref{fig7}. According to the K-S model in Fig.~\ref{fig4}, in the spin-correlated region a doped hole in the underdoped regime of LSCO can itinerate coherently by taking the a$^*_{\rm 1g}$ and b$_{\rm 1g}$ orbitals alternately in the presence of the local AF order without destroying the AF order. 

In the case of a finite spin-correlated region, one may think that there exist the frustrated spins on the boundary between the spin-correlated region of the AF order and the region of the ``resonating valence bond'' (RVB) state proposed by Anderson \cite{Anderson}. Here the frustrated spins mean that the localized spins on the boundary are not in the AF order, but directed parallel to each other. Suppose that one of frustrated spins of a parallel direction on the boundary  has changed  its direction from parallel to antiparallel by the fluctuation effect in the 2D Heisenberg AF spin system  during the time of $\tau_{s}$ defined by $\tau_{s} \equiv \hbar /J$, 
where $J$ is the superexchange interaction ($\sim$0.1 eV). 
In the time of $\tau_{s}$, on the other hand, the hole-carriers at the Fermi level can move with the Fermi velocity inside the spin-correlated region of the AF order. The traveling time of a doped hole at the Fermi level 
over an area of the spin-correlation length 
 is given by $\tau_{\rm F} \equiv \lambda_{s}/v_{\rm F}$,
 where $v_{\rm F}$ is the Fermi velocity of a doped hole at the Fermi level. 
In the case of underdoped LSCO, $\tau_{s}$ is 6 $\times$ $10^{-15}$ sec. 
Since $v_{\rm F}$ is estimated to be 2.4 $\times$ $10^{4}$ m/sec from the dispersion of the \#1 band in Fig.~\ref{fig6},  $\tau_{\rm F}$ is 2 $\times$ $10^{-13}$ sec 
for the underdoped region of $x=0.10$ to $x=0.15$ in LSCO, where for the spin-correlation length $\lambda_{s}$ at $x = 0.15$, we have chosen 50\AA. 
Thus $\tau_{\rm F}$ 
 becomes much longer than $\tau_{s}$. As a result the frustrated spins on the boundary change their directions from parallel to antiparallel before a hole-carrier in the spin-correlated region of the AF order reaches the boundary. Thus a metallic state for a doped hole becomes much wider than the observed spin-correlation length by its passing through the boundary with the mechanism of the K-S model.
 
 \subsection{Occurrence of the AF order in a metallic state to lower the kinetic energy of doped 
 holes}

\begin{figure}
\includegraphics[width=9cm]{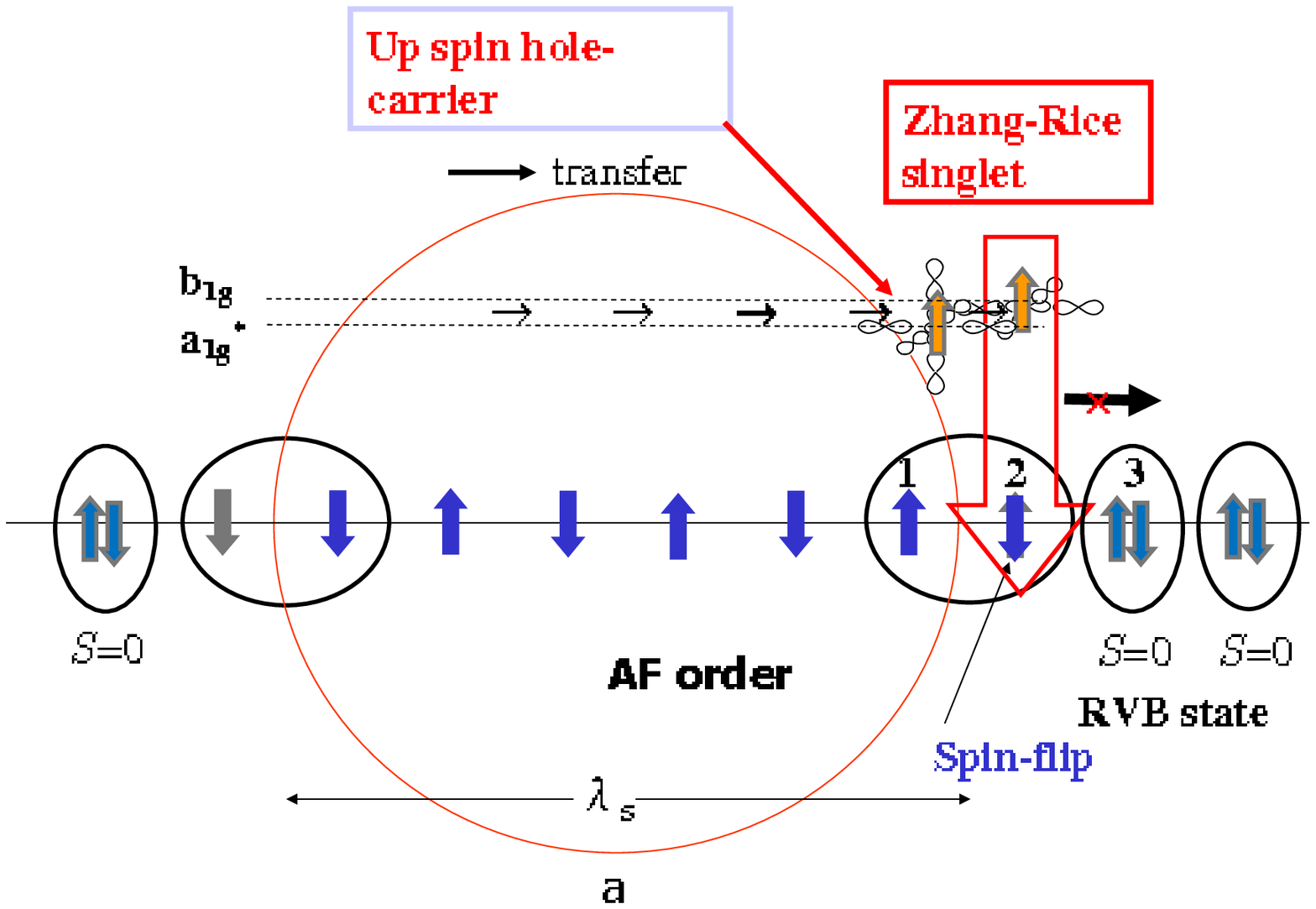}
\includegraphics[width=9cm]{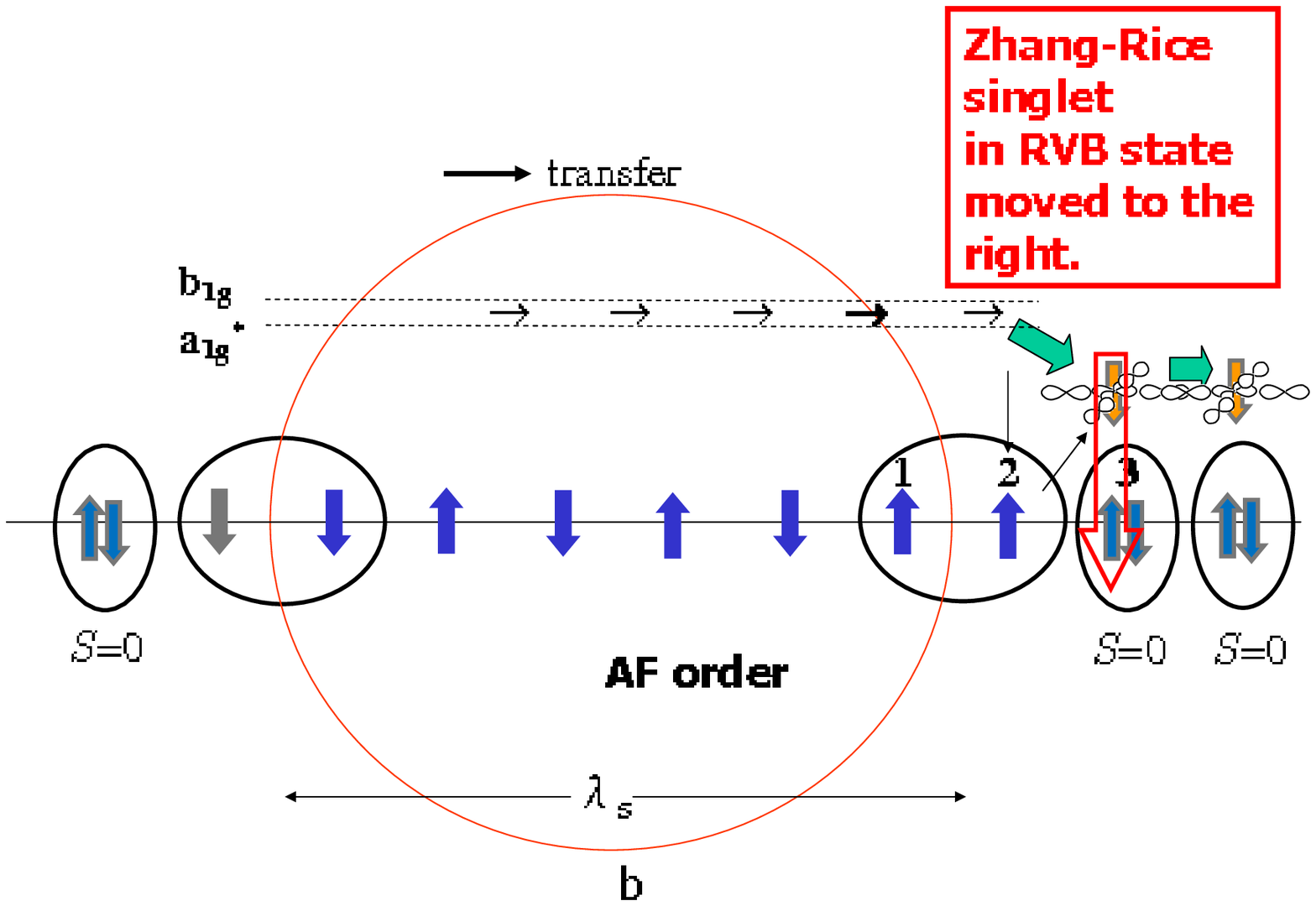}
\caption{\label{fig8}
A special case in which a doped hole reaches site 1 on the boundary and at the same time one of the frustrated spins with up direction at site 2 on the boundary changes its direction from up to down. ({\bf a}) The doped hole with up spin reached site 1 occupies the a$^*_{\rm 1g}$ orbital, where the localized spin at site 2 changes its direction from up to down. Then a doped hole in the a$^*_{\rm 1g}$ orbital at site 1 moves to the b$_{\rm 1g}$ orbital at site 2 on the boundary and it forms Zhang-Rice singlet with the localized down spin at site 2.  ({\bf b}) The Zhang-Rice singlet moves to site 3 in the region of the RVB state like a quasi-particle in the t-J model.}
\end{figure}

  Concerning a finite system of cuprates, Hamada and coworkers \cite{Hamada} and Kamimura and Hamada \cite{KH} tried to determine the ground state of the effective Hamiltonian~(\ref{eq:H}) for the K-S model by carrying out the exact diagonalization of the Hamiltonian ~(\ref{eq:H}) using the Lanczos method for a two-dimensional (2D) square lattice system with 16 (4 $\times$ 4 ) localized spins with the presence of one and two doped holes, respectively. As a result they clarified 
 that, in the presence of hole-carriers, the localized spins in a spin-correlated region tend to form an AF order rather than a random spin-singlet state and 
 thus that the hole-carriers can lower the kinetic energy by itinerating in the lattice of AF order with the mechanism of the K-S model. In this way they have proved that the coexistence of the metallic state and AF order makes the kinetic energy of hole-carriers lower so that such coexistence state corresponds to the lowest energy state in the K-S model. 
Thus generally, a hole-carrier in the spin-correlated region of the AF order can propagate through the boundary of the spin-correlated region with the mechanism of the K-S model and hence a region of a metallic state coexisting with the AF order 
becomes much wider than the observed spin-correlated region.  Recently an idea similar to ours with regard to the lowering of the kinetic energy has been proposed by Wrobel and coworkers who have shown that the lowering of the kinetic energy is the driving mechanism to give rise to superconductivity \cite{Wrobel, Wrobel_Furde}.

As a special case of destroying the coexistence of a metallic state and the local AF order on the boundary, let us consider a case shown in Fig.~\ref{fig8}, where a doped hole with up-spin reaches site 1 on a boundary when one of the frustrated spins with up direction at site 2 is about to change its direction from up to down. Here the localized spins inside the boundary form the AF order while those outside the boundary take the RVB state, as shown in Fig.~\ref{fig8}, where the RVB state consists of the nearest neighbor spin-singlet pairs ($S = 0$), allowing the singlet pairs to move. Thus the doped hole with up spin at site 1 can move to site 2 by occupying the b$_{\rm 1g}$ orbital. As a result it forms the Zhang-Rice singlet with a localized down spin at site 2. In this case the doped hole with up spin in the Zhang-Rice singlet can not move to a$^*_{\rm 1g}$ orbital state at site 3 because the localized spins at site 3 take the RVB state. Thus the coherent motion of a hole-carrier by taking the a$^*_{\rm 1g}$ and b$_{\rm 1g}$ orbitals in the presence of the local AF order without destroying the AF order stops at this moment. 

Instead of it the Zhang-Rice singlet moves from site 2 to site 3 and to further sites in the region of the RVB states like a quasi-particle in the t-J model \cite{Zhang_Rice}, until a certain region of the RVB state changes to the lower energy state of the local AF order to create the coexistence state of a metallic state and the local AF order again. Thus we can say that an spatially inhomogeneous distribution of Fermi-pocket-states and of large-Fermi-surface-states appears as the finite size effect when a temperature is higher than $T_{\rm c}$, and such distribution may change with time. 

Our prediction with regard to the coexistence of Fermi-pocket-states and of large-Fermi-surface-states above $T_{\rm c}$ is consistent with the experimental results by Meng {\it et al} \cite{Meng} who recently reported  the coexistence of Fermi pockets and a large Fermi surface in Bi$_2$Sr$_{2-x}$La$_x$CuO$_{6+\delta}$ (La-Bi2201) which has a similar structure to LSCO with regard to the array of CuO$_{6}$ octahedrons. However, when temperature becomes below $T_{\rm c}$, the regions of the coexistence of a superconducting state and the local AF order becomes dominant, because the energy of such coexistence state is lower than that of the hole-carriers in the RVB state, according to Hamada and his coworkers \cite{Hamada}.

\section{A new phase diagram calculated for underdoped cuprates
 based on the K-S model}

\begin{figure}[h]
\begin{center}
\includegraphics[width=9cm]{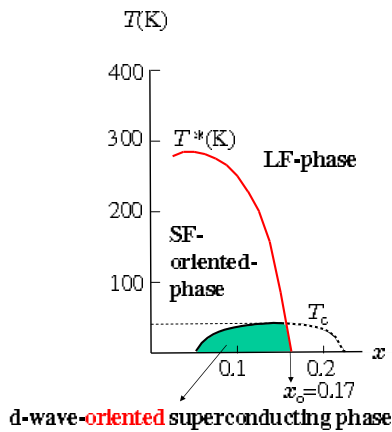}
\end{center}
\caption{\label{fig12}
A new phase diagram for LSCO}
\end{figure}

From the calculated results shown in Fig.~\ref{fig11}{\bf a to c}, we can construct the $T$ vs $x$ phase diagram for LSCO by choosing $T^*(x)$ as a phase boundary, as shown in Fig.~\ref{fig12}. In this phase diagram we have clarified the physical meaning of each area: There is no longer a pseudogap phase. Below $T^*(x)$ in the 
underdoped region the SF phase constructed from the Fermi pockets appears under the coexistence of the local AF order. However, owing to various effects such as the mixing effects of the SF and LF phases due to temperature, hole-concentration and the finite size effect of a metallic state, each new phase is not sharply defined so that we have used a word ``oriented''. For example, we 
have pointed out a possibility of the spatially inhomogeneous distribution of Fermi-pocket-states and of large-Fermi-surface-states above $T_{\rm c}$. In such a way the LF-phase may be mixed into the SF-phase spatially and/or thermally even below $T^*(x)$. Thus, when temperature increases at a certain hole concentration in the underdoped region, the shape of a Fermi pocket changes to a large Fermi surface gradually with increasing temperature. This fact can explain the strange temperature-evolution of a Fermi arc observed by Norman {\it et al} \cite{Norman_Nature} and Kanigel {\it et al} \cite{Kanigel} without introducing the pseudogap. Further, in the 
superconducting phase below $T_{\rm c}$ indicated by the green color in Fig.~\ref{fig12}, an s-wave component of superconductivity originated from the LF-phase may be mixed into the d-wave superconductivity. Thus we call that ares "d-wave oriented superconducting phase". Such mixing effect was experimentally reported by M\"{u}ller \cite{Mueller}. Thus it is better to say that $T^*(x)$ represents a crossover from the SF-phase to the LF-phase rather than a phase boundary. 

Further we can predict that the spin susceptibility shows the 2D-like antiferromagnetic features mainly below $T^*(x)$ while the Pauli-like temperature-dependent behavior above $T^*(x)$. We find that this prediction is also consistent with the experimental results for LSCO \cite{Cooper, Nakano}. In this context it should be emphasized that the K-S model is shown to explain successfully not only the ARPES experimental results \cite{Meng, Tanaka, Norman_Phys_Rev, Lee} but also a number of other experimental results such as NMR results showing the coexistence of a superconducting state and antiferromagnetic order \cite{Mukuda}, polarized X-ray absorption spectra \cite{Chen, Pellegrin}, site-specific X-ray absorption spectroscopy \cite{Merz_Nucker}, anomalous electronic entropy \cite{Loram_Entropy, Kamimura_Entropy}, d-wave superconductivity \cite{Wollman_d-Wave, Tsuei}, etc., without introducing adjustable parameters.  Theoretically the K-S model is also supported by {LDA + U} band calculations \cite{Anisimov}, as already mentioned in Section IIE. 

\section{Conclusion and concluding remarks}

In this paper we have shown on the basis of the K-S model how the interplay of Mott physics and Jahn-Teller physics plays an important role in determining the superconducting as well as the metallic state of underdoped cuprates. It was pointed out for underdoped cuprates that Mott physics gives rises to the existence of local antiferromagnetic order due to the localized spins while that the anti-Jahn-Teller effect as a central issue of Jahn-Teller physics creates the existence of two kinds of orbitals parallel and perpendicular to a CuO$_2$ plane which are energetically nearby. The K-S model which bears important characteristics born from the interplay of Jahn-Teller Physics and Mott Physics has led to the 
coexistence of the local AF order and a metallic states above $T_{\rm c}$. This coexistence has resulted in the occurrence of Fermi pockets. Further below $T_{\rm c}$ the superconductivity and antiferromagnetism coexist, leading to the appearance of d-wave superconductivity even in the phonon-involved mechanism, as was shown by Kamimura, Matsuno, Suwa and Ushio \cite{Kamimura_d-Wave}. 

In connection with the interplay of Jahn-Teller physics and Mott physics, the following important results have been obtained in this paper: It has been clarified on the basis of the K-S model that a concept of pseudogap discussed theoretically \cite{Schmalien, Yang_Rice_Zhang, Wrobel} and reported by ARPES, STM and tunneling experiments below $T^*(x)$ in underdoped cuprates \cite{Norman_PRB2007, Kanigel, Renner} is not necessary. We have shown that the strange phenomena observed in the antinodal region are explained by the real transitions of photo-excited electrons from the occupied states in the highest conduction band in the antinodal region to a free-electron state above the vacuum level. In this context we conclude that the concept of the pseudogap in the underdoped cuprates is no longer necessary. Further the physical meaning of the $T^*$ is not related to the pseudogap but it represents a crossover line from the phase consisting mainly of the Fermi pockets in the normal state to the phase consisting of a large Fermi surface. 

Finally several remarks are made on the small Fermi surface and the shadow bands in the underdoped regime of cuprates. In 1996 Wen and Lee developed a slave-boson theory for the t-J model at finite doping, and showed that Fermi pockets at low doping continuously evolved into the large Fermi surface at high doping concentrations \cite{Xiao-Gang Wen}. Although their theoretical model is different from the K-S model, it is interesting to find that they obtained a similar result to the prediction by the K-S model in 1994 with regard to the change from small FS to a large FS with increasing the hole concentration. Recently a proposal was made to reconcile the experimental result of the coexistence of antiferromagnetism and superconductivity \cite{Kaul}. Further, in relation to the small Fermi surface the idea of a shadow Fermi surface was proposed as a replica of the main Fermi surface transferred by $Q = (\pi/a, \pi/a)$ by Kampf and Schrieffer theoretically \cite{Kampf} and then by Aebi {\it et al} experimentally \cite{Aebi}. Checking the idea of the shadow FS experimentally, the observation of shadow bands
 in the ARPES spectra have been reported \cite{Osterwalder, Vilk, Saini, Nakayama}. Responding to the problems of shadow FS and shadow bands from the standpoint of the K-S model it should be emphasized that the Fermi pockets in the metallic state calculated from the K-S model have been derived by the two-component theory as the result of the interplay of Jahn-Teller physics and Mott physics and thus the origin of Fermi pockets is different from that of a single-component theory.
 Therefore, the Fermi pockets shown in Fig.~\ref{fig7} are neither the shadow Fermi surface nor related to the shadow bands. Thus we conclude that the Fermi pockets in the present paper belong to a new category of a small Fermi surface derived from the interplay of Jahn-Teller physics and Mott physics. 

\begin{acknowledgments}
We would like to thank Dr. Wei-Shen Lee, Prof. Atsushi Fujimori and Prof. Tomohiko Saitoh for their valuable discussion on experimental results and Dr. Jaw-Shen Tsai for valuable comments on the present work. This work was supported by Quantum Bio-Informatics Center in Tokyo University of Science.
\end{acknowledgments}


\end{document}